\documentclass{fac}
\usepackage{graphicx}
\usepackage{amssymb}
\usepackage{amsfonts}
\usepackage{amsmath}
\usepackage{dsfont}

\ifprodtf \else \usepackage{latexsym}\fi

\newcommand\black{\ensuremath{\blacktriangleright}}
\newcommand\white{\ensuremath{\vartriangleright}}

\newif\ifamsfontsloaded
\ifprodtf
  \newcommand\whbl{\white\kern-.1em--\kern-.1em\black}
  \newcommand\blwh{\black\kern-.1em--\kern-.1em\white}
  \newcommand\blbl{\black\kern-.1em--\kern-.1em\black}
  \newcommand\whwh{\white\kern-.1em--\kern-.1em\white}
  \amsfontsloadedtrue
\else
  \checkfont{msam10}
  \iffontfound
    \IfFileExists{amssymb.sty}
      {\usepackage{amssymb}\amsfontsloadedtrue
       \newcommand\whbl{\white\kern-.125em--\kern-.125em\black}%
       \newcommand\blwh{\black\kern-.125em--\kern-.125em\white}%
       \newcommand\blbl{\black\kern-.125em--\kern-.125em\black}%
       \newcommand\whwh{\white\kern-.125em--\kern-.125em\white}}
      {}
  \fi
\fi

\title[A Non-Cooperative Game Model for Reliability-Based Task Scheduling in Cloud Computing]
      {A Non-Cooperative Game Model for Reliability-Based Task Scheduling in Cloud Computing}

\author[Yong Wang]
    {Kai Li\\
    Yong Wang\\
    Meilin Liu\\
     School of Computer Science and Technology,\\
     Beijing University of Technology, Beijing, China\\
     }

\correspond{Yong Wang, Pingleyuan 100, Chaoyang District, Beijing, China.
            e-mail: wangy@bjut.edu.cn}

\pubyear{2011}
\pagerange{\pageref{firstpage}--\pageref{lastpage}}

\begin{document}
\label{firstpage}

\makecorrespond

\maketitle

\begin{abstract}
Cloud computing is a newly emerging distributed system which is evolved from Grid computing. Task scheduling is the core research of cloud computing which studies how to allocate the tasks among the physical nodes, so that the tasks can get a balanced allocation or each task's execution cost decreases to the minimum, or the overall system performance is optimal. Unlike task scheduling based on time or cost before, aiming at the special reliability requirements in cloud computing, we propose a non-cooperative game model for reliability-based task scheduling approach. This model takes the steady-state availability that computing nodes provide as the target, takes the task slicing strategy of the schedulers as the game strategy, then finds the Nash equilibrium solution. And also, we design a task scheduling algorithm based on this model. The experiments can be seen that our task scheduling algorithm is better than the so-called balanced scheduling algorithm.
\end{abstract}

\begin{keywords}
Cloud Computing, Task Scheduling, Steady-State Availability, Non-Cooperative Game, Task Slicing Strategy, Nash Equilibrium Solution.
\end{keywords}

\section{Introduction}

As an Internet-based computing model of public participation, cloud computing\cite{1} is a large-scale distributed computing paradigm that is driven by economies of scale, in which a pool of abstracted, virtualized, dynamically-scalable, managed computing power, storage, platforms, and services are delivered on demand to external customers over the Internet. With the rapid development of the Internet, real-time data stream and connected devices' diverse development, SOA's adoption as well as the promotion of search service, social networks, mobile commerce and open collaboration's demands, cloud computing develops rapidly. Currently, Google, IBM, Amazon, Microsoft and other IT vendors are making great efforts to research and promote cloud computing.

The characteristics of cloud computing, such as abstract and virtualized resources, on demand request of the Internet users, isolation for the accesses of different users, dynamic scalability, security under the unsafe networks, and also the special requirements for reliability because of the public infrastructure role acted by a cloud usually, determine that cloud computing has its own technical architecture as a new distributed computing paradigm. These characteristics will greatly influence the technologies in cloud computing corresponding to those in traditional distributed computing and distributed systems, such as task scheduling technology.

Cloud computing not only provides various business applications or personal applications via the Internet, but also includes the integration of tradition high performance computing under the new environments, for example, Amazon's elastic compute cloud provides high performance computing for the Internet users. Under the large-scale demands for services, especially for the requirements of computing intensive services, how to allocate resources effectively and reliably, and how to handle service requests is particularly important for cloud computing platform. There is no doubt that each characteristics of cloud computing will make the task scheduling in cloud computing special, such as the special requirements for reliability. Currently, almost numerous research works on task scheduling in distributed systems and several works on task scheduling in clouds mostly take time, QoS or the execution cost as the goal without taking into account the reliability of the service provided.

Aiming at the special requirements for reliability in a cloud, according to the nature and characteristics of powerfully parallel processing capabilities for cloud computing, in this paper, we introduce a reliability based task scheduling approach. This approach satisfies the special requirements for reliability in cloud computing to some extent and is also novel in traditional distributed computing. We use the game theory as the main research tool to study the task scheduling problem based on the reliability, namely the steady-state availability that each computing node provides. And also we design a task scheduling algorithm by analyzing the game phenomenon between the cloud computing system components. This scheduling method uses reliability as the accordance of scheduling decision, and is suitable for cloud computing platforms.

In Section $2$, we introduce the related works briefly. We discuss the system structure of a cloud, and introduce a cloud computing system model for computing intensive requirements in
Section $3$. In Section $4$, we establish the non-cooperative game model based on steady-state availability and design the task scheduling algorithm. In Section $5$, we do a number of detailed experiments which show that this scheduling algorithm is better than the so-called balanced scheduling algorithm. Finally, we conclude this paper in Section 6.

\section{Related Works}

In general, job allocation algorithms in distributed systems can be classified as static or dynamic \cite{2}. In static algorithms, job allocation decisions are made at compile time and remain constant during runtime. For example, in \cite{3}, Kim and Kameda proposed a simplified load balancing algorithm, which targets at the minimizing the overall mean job response time via adjusting the each node's load in a distributed computer system that consists of heterogeneous hosts, based on the single-point algorithm originally presented by Tantawi and Towsley. Grosu and Leung\cite{4} formulated a static load balancing problem in single class job distributed systems from the aspect of cooperative game among computers. Also, there exists several studies on static load balancing in multi-class job systems \cite{5}\cite{6}. In contrast, dynamic job allocation algorithms attempt to use the runtime state information to make more informative job allocation decisions. In \cite{7}, Delavar introduced a new scheduling algorithm for optimal scheduling of heterogeneous tasks on heterogeneous sources, according to Genetic Algorithm which can reach to better makespan and more efficiency. In \cite{8}, Fujimoto proposed a new algorithm RR that uses the criterion called total processor cycle consummation, which is the total number of instructions the grid could compute until the completion time of the schedule, regardless how the speed of each processor varies over time, the consumed computing power can be limited within $(1+m(\ln(m-1)+1)/n)$( $m$ represents the number of the processor, n represents the number of independent coarse-grained tasks with the same length)times the optimal one.

For balanced task scheduling, \cite{9,10,11} proposed some models and task scheduling algorithms in distributed system with the market model and game theory. \cite{12,13} introduced a balanced grid task scheduling model based on non-cooperative game. QoS-based grid job allocation problem is modeled as a cooperative game and the structure of the Nash bargaining solution is given in \cite{2}. In \cite{14}, Wei and Vasilakos presented a game theoretic method to schedule dependent computational cloud computing services with time and cost constrained, in which the tasks are divided into subtasks. The above works generally take the scheduler or job manager as the participant of the game, take the total execution time of tasks as the game optimization goals and give the proof of the existence of the Nash equilibrium solution and the solving Nash equilibrium solution algorithm, or model the task scheduling problem as a cooperative game and give the structure of the cooperative game solution.

In a cloud computing environment, the goal of task scheduling is to achieve the optimal scheduling of jobs submitted by users, and try to improve the overall throughput of the cloud computing system. In recent years, a lot of people have been studying the task scheduling problems in the cloud computing environment and made rich achievements. At present, task scheduling algorithms at home and broad mainly base on the earliest completion time ,quality of service, load balancing, economic principles and so on \cite{15}. Job allocation algorithms can be classified as performance-centric, QoS-centric and economic principle-centric based on the different goals. Performance-centric task scheduling algorithms take the scheduling performance as the ultimate goal such as the shortest completion time, including Min-Min, Max-Min algorithm (e.g.,\cite{16,17,18}), genetic algorithm (e.g.,\cite{19}), ant colony algorithm (e.g.,\cite{20,21}). QoS-centric task scheduling algorithms have been studied widely. In \cite{22}, He and Sun improved Min-Min algorithm, which optimizes the system throughput according to whether the user has QoS requirements or not. In \cite{23}, Chanhan and Joshi selected resources based on the weighted average execution time, regarded the network bandwidth as QoS attributes and divided the tasks into two categories , high QoS and low QoS, and high QoS tasks had the priority to be scheduled. In \cite{24}, Xu and Wang developed a scheduling strategy for multiple workflows with different QoS requirements. In addition, \cite{25,26,27} proposed some task scheduling algorithms from the view of economic principle.

Above research works mostly take time, QoS or the execution cost as the goal without taking into account the characteristics of cloud computing, especially the reliability of the service provided. Unlike the above research, this paper proposes a non-cooperative game model for reliability-based task scheduling in cloud computing system. This model takes the steady-state availability that computing nodes provides as the target, takes the task's slicing strategy in computing nodes as the game strategy, then analyzes the existence of the non-cooperative game Nash equilibrium solutions.

According to the classification of static algorithms or dynamic ones for task scheduling algorithms, our algorithm can be classified into a semi-dynamic algorithm, just because this algorithm utilizes the states and capacity of computing nodes in a statistic way. In the view of QoS-driven task scheduling algorithms, our algorithm provides a new approach based on reliability.

This paper has two main contributions as follows:

\begin{enumerate}
  \item To satisfy the special requirements for reliability in a cloud system, a reliability-based task scheduling model is established. This work is also novel in task scheduling of distributed systems.
  \item Through taking the steady-state availability that computing nodes provide as the goal, we establish a non-cooperative game model on task scheduling in cloud computing system, and give the proof of the existence of the Nash equilibrium solutions, then get the task scheduling algorithm. From the experiments, we can see that this scheduling algorithm has better optimal results than the balanced scheduling algorithm.
\end{enumerate}

\section{System Model for A Cloud System}

\begin{figure}
 \begin{center}
  \includegraphics[width=8cm,height=8cm]{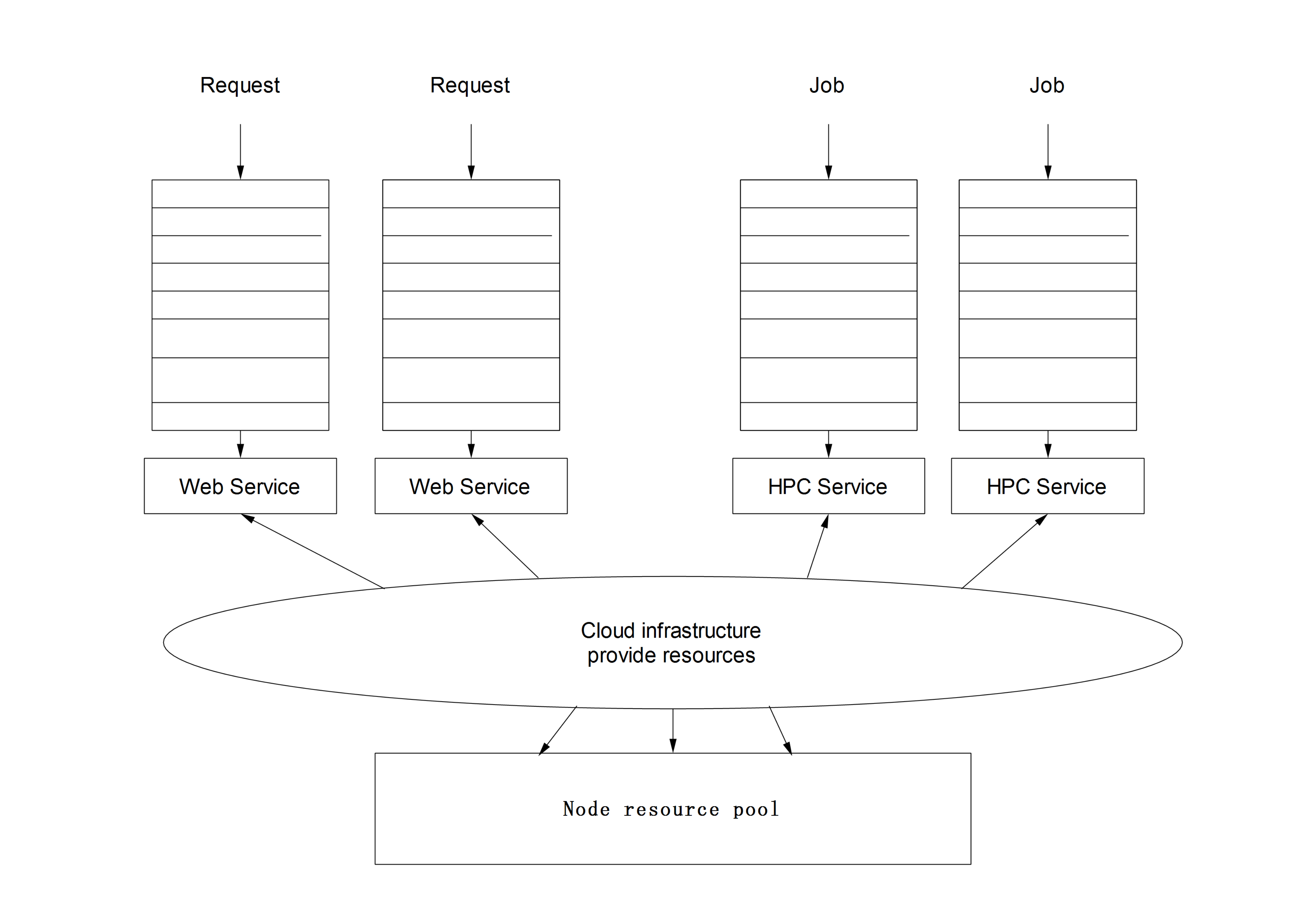}
 \caption{Cloud System Simulation}
 \label{Fig.1}
 \end{center}
\end{figure}

A variety of services in the cloud computing platform can be roughly grouped into two categories: data-compute-intensive services and interaction-intensive services \cite{29}. The former have a higher complexity and requires a higher computing power; the latter can be classified as general Web Services which needs higher real-time requirements. Fig.\ref{Fig.1} is a cloud system simulation based on the two type of services. Firstly, the service request will be classified to the Web Service queue or HPC (High performance computing) service queue according to its type, and then calls cloud resources depending on the type of service.

\begin{figure}
 \begin{center}
  \includegraphics[width=8cm,height=8cm]{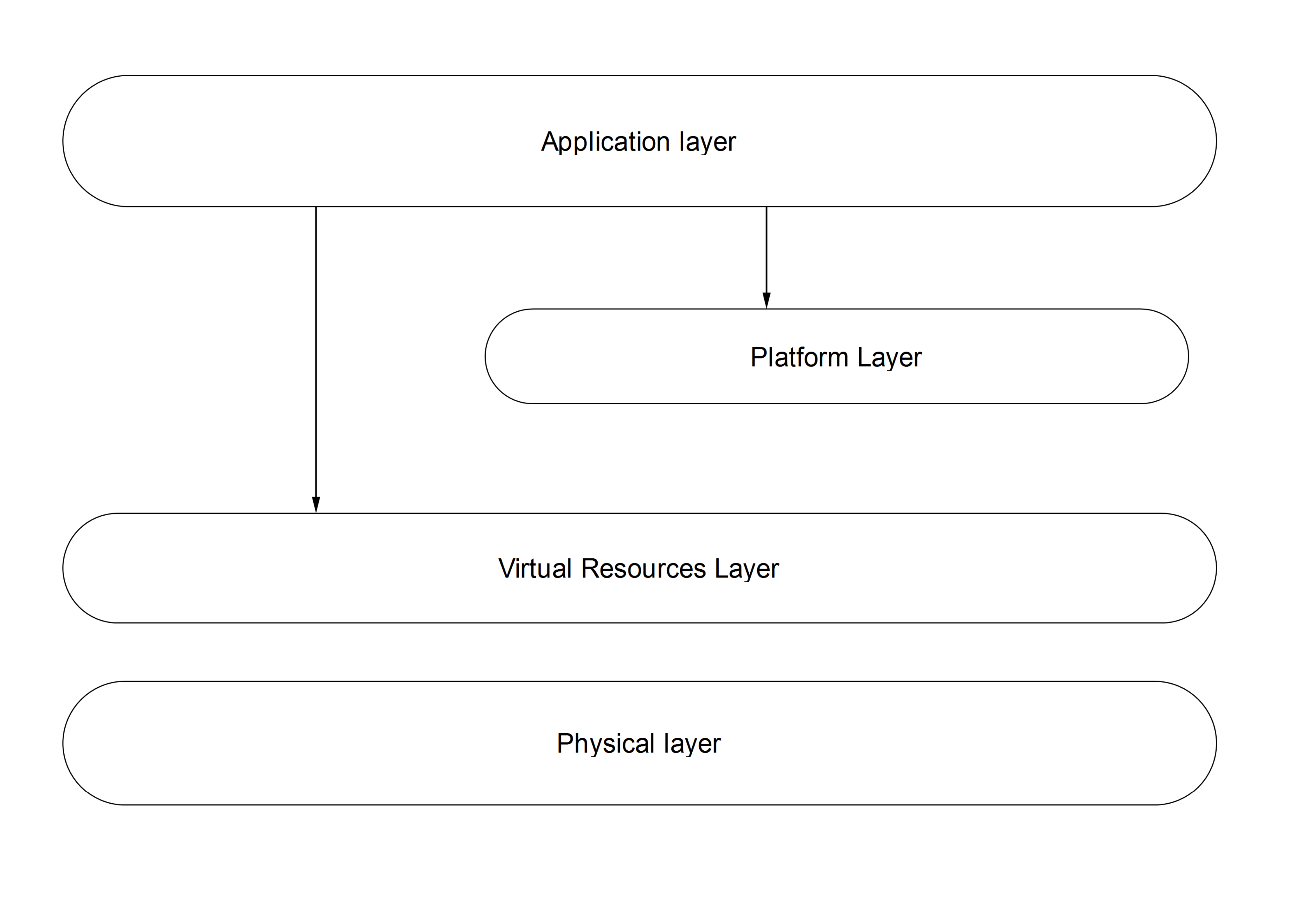}
 \caption{Cloud System Architecture}
 \label{Fig.2}
 \end{center}
\end{figure}

It is well known that the architecture of a cloud system roughly includes application layer, platform layer, virtual resource layer and physical layer \cite{30} shown as Fig.\ref{Fig.2}. The application layer accepts the requests of the Internet users through various applications, then deliver the requests to the platform layer. The platform layer usually contains a task scheduler to split the user tasks into pieces and deliver these task pieces onto different virtual resources. There are mappings from virtual resources to physical resources using the so-called virtualization technology.

In addition, data-compute-intensive services are usually very complex, and for these service requests, a cloud system should give full play to their own advantages of task parallel processing and the following three assumptions should be reasonable:

\begin{enumerate}
  \item In the cloud computing system model, the scheduler can split the task into task slices. Because the configuration of a node is controllable within a mesh range, we can assume that all the computing nodes can handle the task slices;
  \item The internal processing cost of the scheduler (such as the reliability when the task is split) can be ignored, namely, we can assume that the task slices' execution cost in computing nodes is the key consideration of the task execution cost;
  \item In the cloud computing system, node resources that provide service may be subjected to scheduled backups and unpredictable failures.
\end{enumerate}

\begin{figure}
 \begin{center}
  \includegraphics[width=8cm,height=8cm]{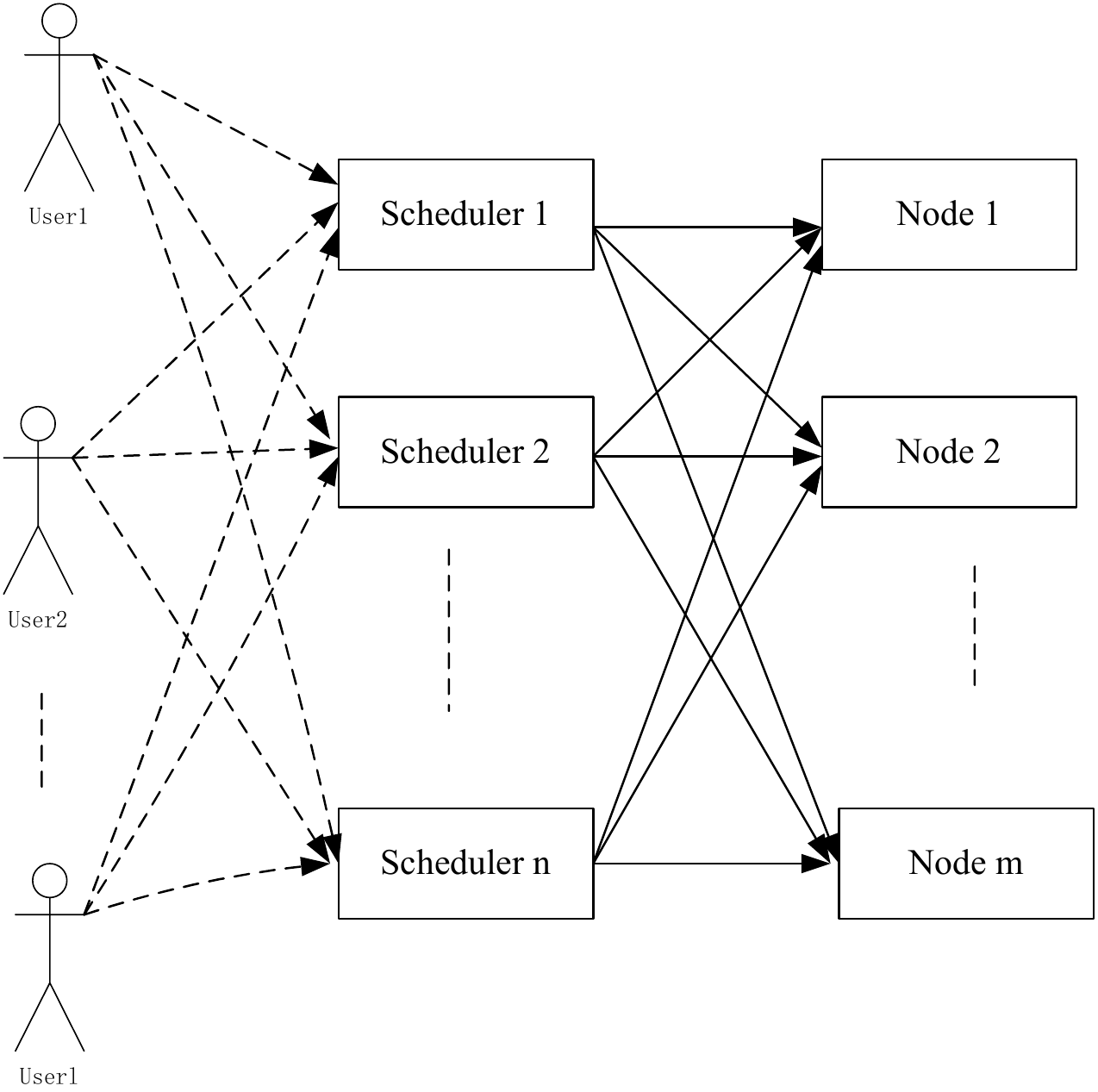}
 \caption{System Model}
 \label{Fig.3}
 \end{center}
\end{figure}

Conjunction with Fig.\ref{Fig.1} and Fig.\ref{Fig.2}, the system model of a cloud system for compute-intensive requirements shown in Fig.\ref{Fig.3} is reasonable. In this system model, the number of computing nodes that can provide service is assumed to be $m$, the number of users that generate service request is $l$, and the number of schedulers that allocates resource to implement the service request is $n$.

\begin{itemize}
  \item User: a user generates a request(task) to a scheduler and user $k$ is assumed to generate jobs with average rate $\beta_k$(jobs per second ) according to a Poisson process independently. Jobs are then sent by the user to a scheduler that dispatches them to the computing nodes.
  \item Scheduler: a scheduler receives job from a set of users and then assigns them to computing nodes in the cloud computing system. According to the first assumption, the task decomposition cost is negligible.
  \item Task slice: depending on the number of computing nodes, scheduler $i$ dispatches the users' tasks into $m$ task slices, $a_{ij}$ is the ratio of scheduler $i$ assigns one task to the computing node $j$ which satisfies the following constraint:
      \begin{equation}
           a_{ij}\geq0 \quad and \quad \sum_{j=1}^m a_{ij}=1
      \end{equation}
  \item Computing node: the computing node executes and processes task slices sent to it. According to the second assumption, the computing node has the ability to execute the general task slice. $\mu_j$is the average processing rate of jobs at computational node $j$, the execution time can be subjected to arbitrary distribution, and each computing node can be modeled as an M/G/1 queuing system with the general retry time and server crashes\footnote[1]{An M/G/1 queuing system with the general retry time and server crashes is a kind of retrial queueing system, which is characterized by the feature that arrivals who find the server unavailable are obliged to leave the service area and to try again for their requests in random order and at random intervals. This type queuing system has been applied widely in computer system and communication networks.}.
      For stability, we have two constrains as follows:
  \begin{equation}
     \sum_{i=1}^n \lambda_i<\sum_{j=1}^m \mu_j
  \end{equation}

  \begin{equation}
     \sum_{i=1}^n \lambda_ia_{ij}<\mu_j
  \end{equation}

Where $\lambda_i$ is the average rate of jobs that scheduler $i$ issues. The first constrain means that the average rate of jobs that all schedulers issues must not be faster than average rate of jobs that all computing nodes execute. The second constrain means that the average rate of jobs sent to computing node $j$ must not exceed the rate at which jobs can be executed by the computing node $j$.
\end{itemize}

\section{Non-Cooperative Game Model for Reliability-Based Task Scheduling}

\subsection{Game Analysis}

Each scheduler which is independent in cloud computing system shares the computing nodes and always expects that steady-state availability that all the computing nodes can provide is the maximum. They compete with each other so that they can constitute a non-cooperative game as participants.

Let $a_i=\{a_{i1},a_{i2},...,a_{im}\}$ represents scheduler $i$'s task slicing scheme on all cloud computing nodes , where $i=1,2,...,n$ and $a_{ij}$ is the task's ratio of jobs according which scheduler $i$ assigns jobs to computing node $j$.

According to reference \cite{28}, for the M/G/1 queuing system with the general retry time and server crashes, the steady-state availability that computing node $j$ can provide is calculated as:

\begin{equation}
    A_j=1-\delta_j\beta_{1j}(1+\mu_j^,\gamma_j)
\end{equation}
 Subject to
\begin{equation}
      0 \leq A_j \leq 1
\end{equation}

Where $A_j$ is the steady-state availability that computing node $j(j=1,2,...,m)$ provides , $\delta_j$ represents the average rate of jobs sent to computing node $j$ according to a Poisson process, $\beta_{1j}$ is the mean time of computing node $j$ serves, $\mu_j^,$ is the average failure rate when the computing node $j$ is busy and $\gamma_j$ is the average retrial time of computing node $j$.

After each scheduler $i(i=1,2,...,n)$ assigns its tasks to computing node $j$ according to $a_{ij}$, the steady-state availability that computing node $j$ provides can be calculated as:
\begin{equation}
      A_j=1-\sum_{i=1}^n(\lambda_ia_{ij})\beta_{1j}(1+\mu_j^,\gamma_j)
\end{equation}

Then take the summary of steady-state availability' reciprocals that all the computing nodes provide to scheduler $i$ as the objective function of game, namely:

\begin{equation}
     D_{i}=\sum_{j=1}^m\frac{1}{1-\sum_{i=1}^n(\lambda_ia_{ij})\beta_{1j}(1+\mu_j^,\gamma_j)}
\end{equation}

In order to facilitate the algorithm description, we introduce a new variable $\mu_{ji}$, shown in (8). $\mu_{ji}$ is defined as computational capacity that computing node $j$ provides to scheduler $i$ which is defined as:

\begin{equation}
    \mu_{ji}=\mu_j-\sum_{k=1,k\neq i}^n\lambda_ka_{kj}
\end{equation}

Substituting (8) into(7),we obtain the objective function as follow:

\begin{equation}
     D_{i}=\sum_{j=1}^m\frac{1}{1-(\lambda_ia_{ij}+\mu_j-\mu_{ji})\beta_{1j}(1+\mu_j^,\gamma_j)}
\end{equation}

\textbf{Theorem $1$}. The above non-cooperative game has only one Nash equilibrium solution and this solution $a_i^*=\{a_{i1}^*,...,a_{ij}^*,...,a_{im}^*\},i=1,2,...,n$ can make formula $7$ minimum.

\begin{proof}
According to reference \cite{11}, it is equivalent to the proof of the formula $8$ for continuous, incremental convex function.
Continuity and  incremental is apparent, so we prove the convexity below, namely the formula $8$ has the first order, second order partial derivatives for each strategy component $a_{ij}$.

$$
\frac{\partial D_{i}}{\partial a_{ij}}=\frac{(1+\mu_j^,\gamma_j)\beta_{1j}\lambda_i}{(1-(\lambda_i a_{ij}+\mu_j-\mu_{ji})\beta_{1j}(1+\mu_j^,\gamma_j))^2}>0
$$

$$
\frac{\partial^2 D_{i}}{\partial a_{ij}^2}=\frac{2(1+\mu_j^,\gamma_j)^2\beta_{1j}^2\lambda_i^2}{(1-(\lambda_ia_{ij}+\mu_j-\mu_{ji})\beta_{1j}(1+\mu_j^,\gamma_j))^3}>0
$$

Thus, the formula $8$ is convex.
\end{proof}

\subsection{Task Scheduling Algorithm}

Theorem $1$ indicates that the formula $8$ is a convex function and its optimal condition meets the Kuhn-Tucker conditions. Our aim is to get the solution than can minimize the objective function.
First, the Lagrangian function is given by

\begin{equation}
     L=\sum_{j=1}^m(\frac{1}{(1-(\lambda_ia_{ij}+\mu_j-\mu_{ji})\beta_{1j}(1+\mu_j^,\gamma_j))}-\alpha *a_{ij})+\alpha
\end{equation}

A necessary condition is

\begin{equation}
     \frac{\partial L}{\partial a_{ij}}=0
\end{equation}

Solving$(11)$, we get the following:

\begin{equation}
   \frac{(1+\mu_j^,\gamma_j)\beta_{1j}\lambda_i}{(1-(\lambda_i a_{ij}+\mu_j-\mu_{ji})\beta_{1j}(1+\mu_j^,\gamma_j))^2}-\alpha=0
\end{equation}

Solving $(12)$ with the constraints $(1),(2),(3)$ and $(5)$,we get the following:

\begin{equation}
     a_{ij}=\frac{1-(1+\mu_j^,\gamma_j)\beta_{1j}(\mu_j-\mu_{ji})-\sqrt{\frac{(1+\mu_j^,\gamma_j)\beta_{1j}\lambda_i}{\alpha}}}{(1+\mu_j^,\gamma_j)\beta_{1j}\lambda_i}
\end{equation}

By the formula $(13)$, we can obtain that when the inequality$(14)$ is satisfied, $a_{ij}<0$, in this case, let $a_{ij}=0$.

\begin{equation}
     \alpha<\frac{(1+\mu_j^,\gamma_j)\beta_{1j}\lambda_i}{(1-(\mu_j-\mu_{ji})\beta_{1j}(1+\mu_j^,\gamma_j))^2}
\end{equation}

Taking into account the compactness of the results, we can get the Nash equilibrium solution of (9) as follows:

\begin{enumerate}
    \item Sort the computing nodes according to the equation below such that $\Theta_{i1}\leq \Theta_{i2}\leq...\leq \Theta_{im}$, where $\Theta_{ij}$ is given by
          \begin{equation}
               \Theta_{ij}=\frac{(1+\mu_j^,\gamma_j)\beta_{1j}\lambda_i}{(1-(\mu_j-\mu_{ji})\beta_{1j}(1+\mu_j^,\gamma_j))^2}
          \end{equation}
    \item If $j>d_i$, then $a_{ij}=0$. $d_i(1\leq d_i\leq m)$ is the maximum positive integer that makes each $a_{ij}$ gotten from solving (16) that satisfies $0\leq a_{ij} \leq 1$ after all computing nodes sorted by $\theta_{ij}$:
    \item Otherwise, calculate $a_{ij}$ with (13), where $\alpha$ is calculated by (17).

\end{enumerate}
\begin{equation}
   \sum_{j=1}^{d_i}\frac{1-(1+\mu_j^,\gamma_j)\beta_{1j}(\mu_j-\mu_{ji})-\sqrt{\frac{(1+\mu_j^,\gamma_j)\beta_{1j}\lambda_i}{\alpha}}}{(1+\mu_j^,\gamma_j)\beta_{1j}\lambda_i}= 1
\end{equation}
\begin{equation}
   \alpha=(\frac{\sum_{j=1}^{d_i}\frac{1}{\sqrt{(1+\mu_j^,\gamma_j)\beta_{1j}}}}{\sum_{j=1}^{d_i}\frac{1-(\mu_j-\mu_{ji})\beta_{1j}(1+\mu_j^,\gamma_j)}{(1+\mu_j^,\gamma_j)\beta_{1j}\lambda_i}})^2\frac{1}{\lambda_i}
\end{equation}
where $j=index(i,j)$ in (16) and (17).

On the basis of the above process, we can design an approximate algorithm on the basis of the above method as the following process: each scheduler calculates a best task slicing scheme $a_{i}$ that results in minimum $D_{i}$ in cycles until all schedulers' minimum values of $D_{i}$ reach an equilibrium. In fact, as can be seen from (7), after the task scheduling algorithm is calculated, $D_{i}$ doesn't vary with $i$, namely the minimum value of all schedulers's objective function value is equal. Therefore, whether the all schedulers' minimum values of $D_{i}$ reach an equilibrium or not can be judged by the difference of $D_{i}$ calculated from each $a$ in cycles.

On the basis of the above analysis, we obtain the task scheduling algorithm as follows:

\begin{itemize}
  \item Step $1$. System parameters initialization: Let $n$ represents the number of schedulers in the cloud computing system, $m$ represents the number of the computing nodes, $\lambda_i(0)$ is the average rate of jobs that scheduler $i$ issues, $\mu_j(0)$ is the average processing rate of jobs at computing node $j$,where $i=1,2,...,n, j=1,2,...,m$. The task slice program of scheduler $i$ is initialized as: $a_i(0)=\{a_{i1}(0),a_{i2}(0),...,a_{im}(0)\}=\{\frac{1}{m},\frac{1}{m},...,\frac{1}{m}\}$, let $\mu_j^,=\frac{\mu_j(0)}{10}, \gamma_j(0)=\frac{5}{\mu_j(0)}$, the error precision of the objective function is $\varepsilon(0)=1$;
  \item Step $2$. Calculate the objective function value $latterD$ under the initial conditions by the formula(7);
  \item Step $3$. If the error precision $\varepsilon\leq 10_{-6}$ then the task scheduling program is calculated and the cycle ends; otherwise, repeat the following step 3.1 to step 3.5;
  \begin{itemize}
    \item Step $3.1$. Let $formerD=latterD,$, $formerD$ is the temporary value of objective function under the previous scheduling program;
    \item Step $3.2$. Calculate $\mu_{ji}$ by the formula $\mu_{ji}=\mu_j-\sum_{k=1,k\neq i}^n\lambda_ka_{kj}$;
    \item Step $3.3$. Calculate $\Theta_{ij}$ by the formula $ \Theta_{ij}=\frac{(1+\mu_j^,\gamma_j)\beta_{1j}\lambda_i}{(1-(\mu_j-\mu_{ji})\beta_{1j}(1+\mu_j^,\gamma_j))^2}$, sort $\Theta_{ij}$ in the ascending order, and store the sorting result into the variable $index$;
    \item Step $3.4$. Perform the following steps from $i=1$ to $n$;
    \begin{itemize}
      \item Step $3.4.1$. Perform the following steps from $d_i=m$ to $1$;
      \begin{itemize}
        \item Step $3.4.1.1$. Calculate $\alpha$ by the formula $\alpha=(\frac{\sum_{j=1}^{d_i}\frac{1}{\sqrt{W_{index(i,j)}}}}{\sum_{j=1}^{d_i}\frac{1-W_{index(i,j)}(\mu_{index(i,j)}-\mu_{index(i,j),i})}{W_{index(i,j)}\lambda_i}-1})^2\frac{1}{\lambda_i}$
            where $W_{index(i,j)}=(1+\mu_{index(i,j)}^,\gamma_{index(i,j)}\beta_{1,index(i,j)}$;
        \item Step $3.4.1.2$. Calculate $a_{i,index(i,j)}$ by the formula $ a_{ij}=\frac{1-(1+\mu_j^,\gamma_j)\beta_{1j}(\mu_j-\mu_{ji})-\sqrt{\frac{(1+\mu_j^,\gamma_j)\beta_{1j}\lambda_i}{\alpha}}}{(1+\mu_j^,\gamma_j)\beta_{1j}\lambda_i}$,
            where $j=index(i,j)$, $a_{i,index(i,j)}$ is used to determine the task  proportion that scheduler $i$ assigns to the computing node $index(i,j)$;
        \item Step $3.4.1.3$. if $a_{ij}$ that doesn't satisfy $0\leq a_{ij}\leq 1$,then let $d_i=d_i-1$,return to step 3.4.1.1; otherwise, go on;
      \end{itemize}
      \item Step $3.4.2$. Modify $\mu_{ji}$ using obtained $a$;
      \item Step $3.4.3$. Calculate $\Theta_{ij}$ by the formula $\Theta_{ij}=\frac{(1+\mu_j^,\gamma_j)\beta_{1j}\lambda_i}{(1-(\mu_j-\mu_{ji})\beta_{1j}(1+\mu_j^,\gamma_j))^2}$, sort $\Theta_{ij}$ in the ascending order, and store the sorting result into the variable $index$;
      \item Step $3.4.4$. Let $i=i+1$,return to step $3.4.1$;
    \end{itemize}
    \item Step $3.5$. Calculate the objective function value $latterD$ using the revised $a$ by the formula $7$, $\varepsilon=|formerD-latterD|$, return to step $3$.
  \end{itemize}
\end{itemize}

The above task scheduling algorithm determines the task slicing program according to which each scheduler should allocate jobs to each computing node in order to minimize the reliability level of all the computing nodes.

\section{Experiments}

In order to judge the effects of the above algorithm, we intend to use Balanced scheduling algorithm\cite{32} for comparison, which is labeled as BSA. The algorithm this paper proposes is labeled as RBSA. The scheduling strategy of BSA is calculated under the below formula:

\begin{equation}
     a_{ij}=\frac{\mu_{ji}}{\sum_{j=1}^m\mu_{ji}}
\end{equation}

Let $\phi_i$ represents the relative job arrival rate of scheduler $i$, the average arrival rate of scheduler $i$ is calculated under the below formula:

\begin{equation}
     \lambda_i=\phi_i*\rho*\sum_{j=1}^m\mu_j
\end{equation}

Where $\rho$ is the required overall average system load.
Here we do some experiments from five aspects as follows.

\subsection{Objective Function Value}\label{OFV}

Considering that the processing abilities of computing nodes in the cloud computing system are balanced or not, we do two sets of experiments independently.
In the first set of experiment, there are some computing nodes that have higher average processing rate than the others clearly. We set the average system load to $50$ percent($\rho=0.5$), the number of the scheduler is $10$ and the number o f the computing node is $15$. The relative job arrival rate of each scheduler is listed at Table \ref{table.1}.

  \begin{center}
  \begin{table}
  \caption{Relative Job Arrival Rate of Each Scheduler}
  \label{table.1}
  \begin{tabular}{|c|c|c|c|}
    \hline
    scheduler & 1 & 2-5 & 6 \\
    \hline
    Relative job arrival rate & 0.0035 & 0.01 & 0.006  \\
    \hline
    scheduler & 7 & 8 & 9-10 \\
    \hline
    Relative job arrival rate & 0.005 & 0.002 & 0.001 \\
    \hline
  \end{tabular}
  \end{table}
  \end{center}

Table \ref{table.2} shows the average processing rate of the computing nodes in the cloud computing system.

  \begin{center}
  \begin{table}
  \caption{Average Task Processing Rate of the Computing Nodes}
  \label{table.2}
  \begin{tabular}{|c|c|c|c|}
    \hline
    Computing node & 1-7 & 8-10 & 11\\
    \hline
    Average Task Processing Rate & 0.02 & 0.033 & 0.0231 \\
    \hline
    Computing node  & 12 & 13 & 14\\
    \hline
    Average Task Processing Rate  & 0.02511 & 0.0153 & 0.023 \\
    \hline
    Computing node  & 15 & & \\
    \hline
    Average Task Processing Rate  & 0.025 & & \\
    \hline
  \end{tabular}
  \end{table}
  \end{center}

\begin{figure}
 \begin{center}
  \includegraphics[width=8cm,height=8cm]{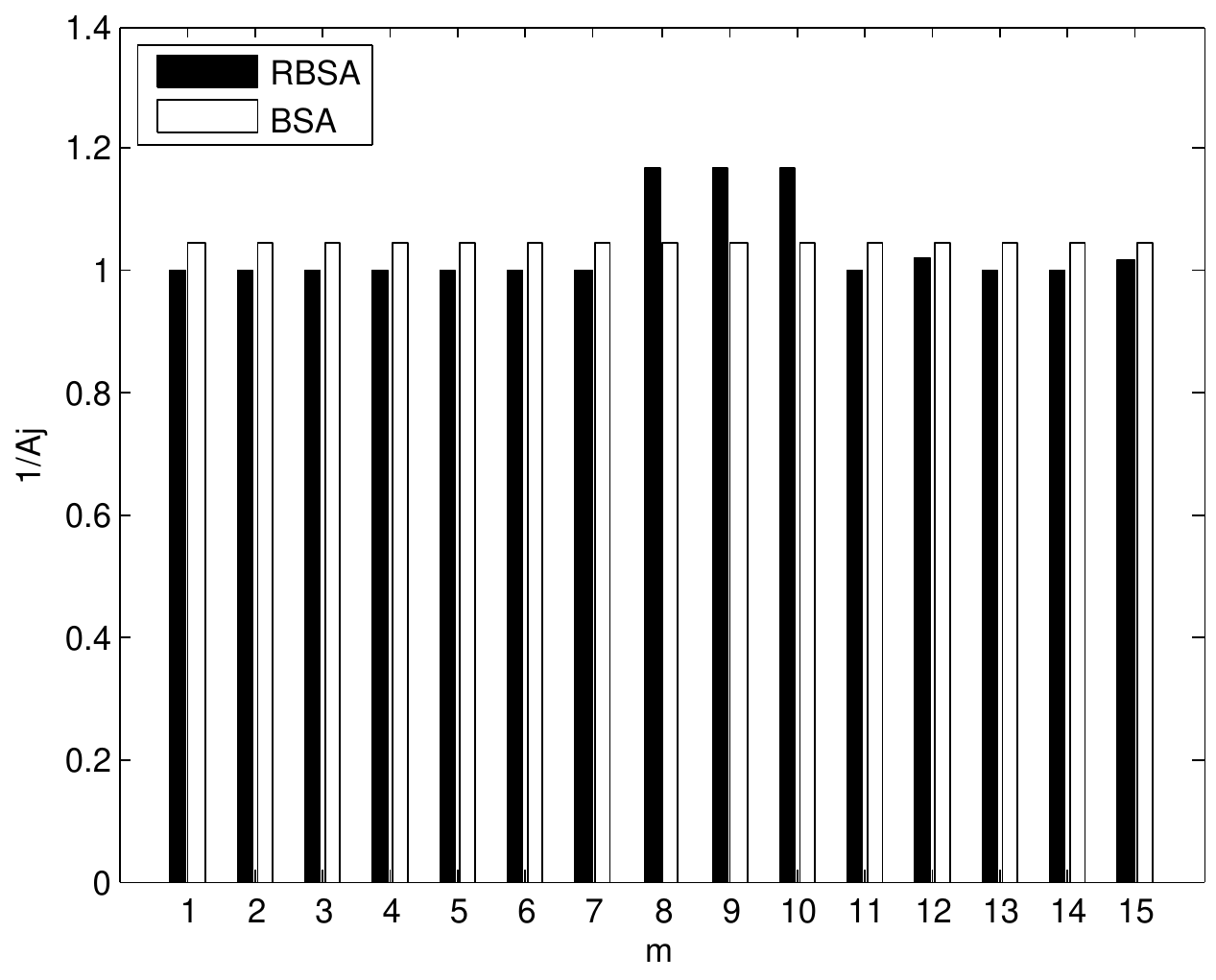}
 \caption{Reciprocal of Steady-State Availability Each Computing Node Provides}
 \label{Fig.4}
 \end{center}
\end{figure}

Under the above initial condition, we can get the reciprocal of the steady-state availability that each computing node provides separately by RBSA and BSA as Fig \ref{Fig.4} shows.

As can be seen from the Fig.\ref{Fig.4}, the BSA scheme is the scheme that just results in each computing node having the same processing ability; the RBSA assigns appropriate jobs to each computing node, namely, the computing nodes that have higher processing ability are assigned more jobs so that the system as a whole can provide the maximal processing ability to the users.
By summing the reciprocal values of RBSA and BSA, we find that the summary from RBSA is $0.0139$ smaller than BSA, namely, RBSA can make the system provide higher steady-state availability than BSA.

In the second set of experiments, the processing ability of each node provides is approximate and the remaining conditions are the same as the first experiment. The average processing rates of the computing nodes are shown in Table \ref{table.3}.

\begin{center}
  \begin{table}
  \caption{Average Task Processing Rate of the Computing Nodes}
  \label{table.3}
    \begin{tabular}{|c|c|c|c|c|}
    \hline
    Computing node & 1 & 2 & 3-4 & 5 \\
    \hline
    Average Task Processing Rate & 0.031 & 0.03 & 0.029 & 0.031\\
    \hline
    Computing node & 6-7 & 8-10 & 11 & 12 \\
    \hline
    Average Task Processing Rate & 0.03 & 0.033 & 0.028 & 0.029 \\
    \hline
    Computing node & 13-14 & 15 & & \\
    \hline
    Average Task Processing Rate & 0.030 &0.031 & & \\
    \hline
  \end{tabular}
  \end{table}
\end{center}

\begin{figure}
 \begin{center}
  \includegraphics[width=8cm,height=8cm]{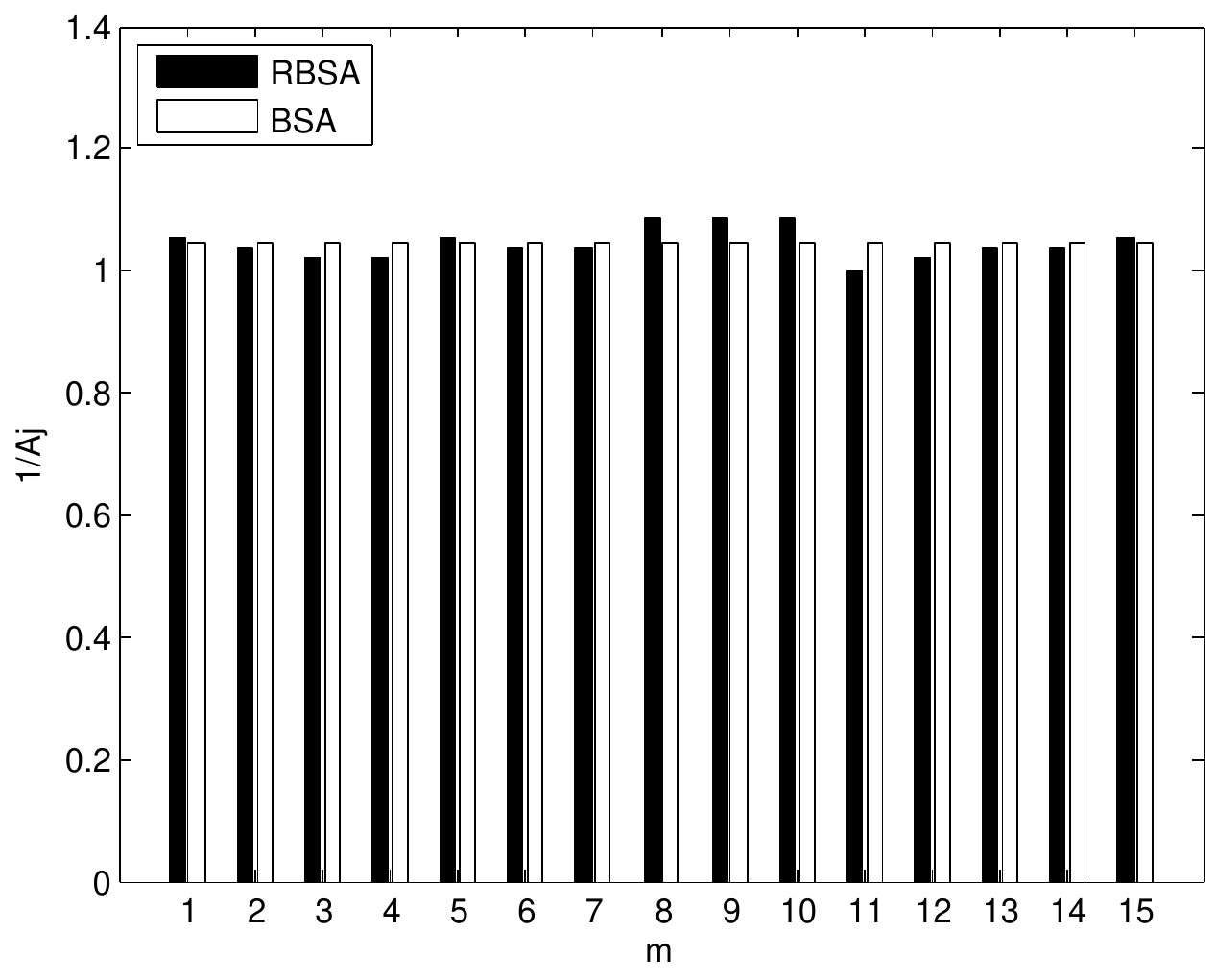}
 \caption{Reciprocal of Steady-state Availability Each Computing Node Provides}
 \label{Fig.5}
 \end{center}
\end{figure}

Under the above initial condition, we can get the reciprocal of the steady-state availability that each computing node provides separately by RBSA and BSA as Fig.\ref{Fig.5} shows.

By summing the reciprocal values of RBSA and BSA, we find that the summary from RBSA is $0.0094$ smaller than the summary of BSA, namely, RBSA can also make the system provide higher steady-state availability than BSA.

According to the above two sets of experiments, we can draw this conclusion: whether the processing ability is balanced or not, the task scheduling algorithm this paper proposes is better than the Balanced scheduling algorithm.

\subsection{Effect of System Load}

In the second part of the experiment, we consider the effect of the rate of arrival tasks. Here, we vary the average load of the system from $10$ percent to $90$ percent, the remaining conditions are the same as the first experiment in Section \ref{OFV}.

\begin{figure}
 \begin{center}
  \includegraphics[width=8cm,height=8cm]{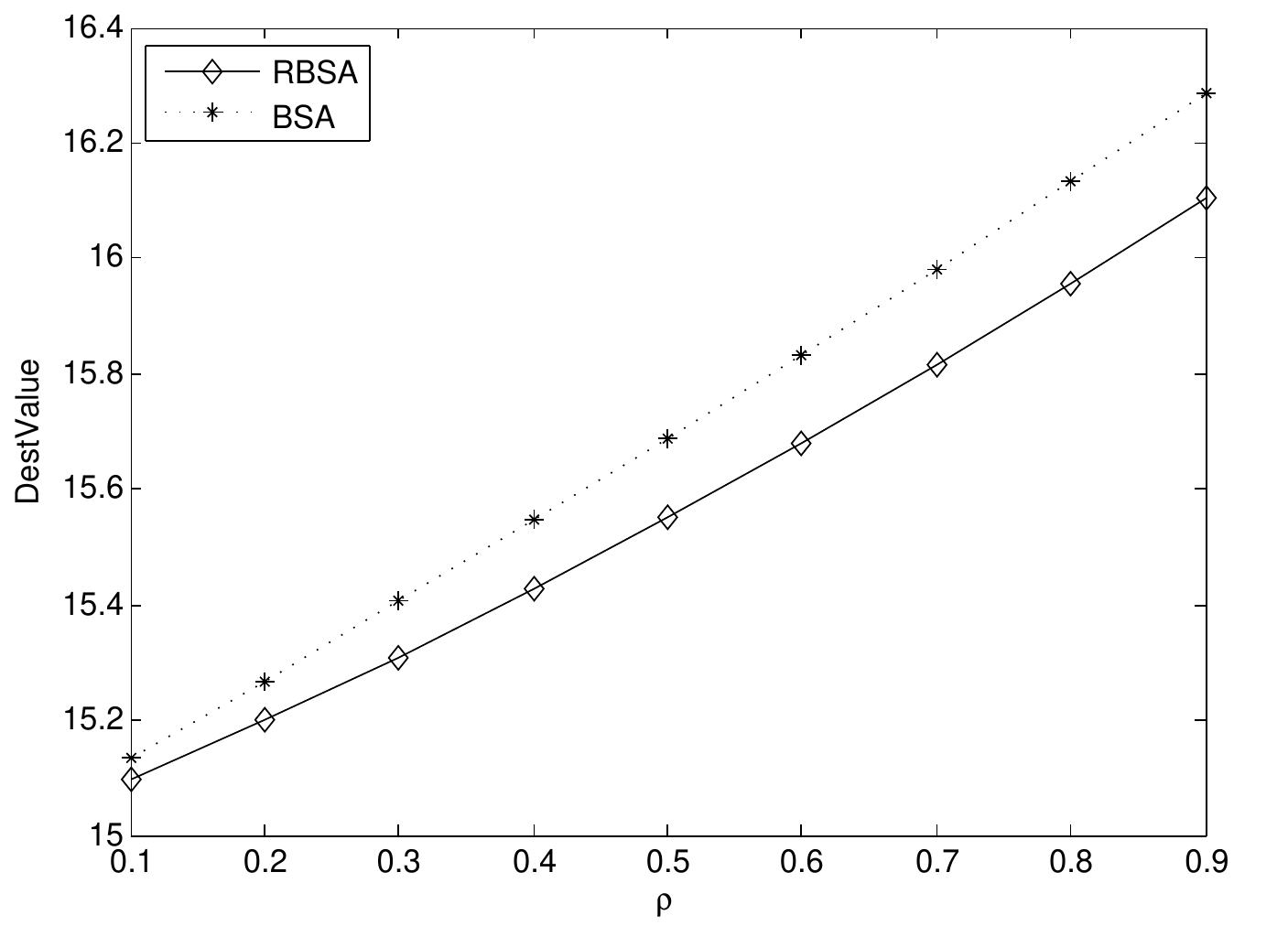}
 \caption{Objective Function Value Versus System Load}
 \label{Fig.6}
 \end{center}
\end{figure}

Under the above initial condition, we can get the objective function values under different system load separately by RBSA and BSA as Fig.\ref{Fig.6} shows.

Fig.\ref{Fig.6} shows the objective function values as the system load is varied from $10$ percent to $90$ percent. Comparing the RBSA and BSA schemes, we can see that both increase while the system load increases. What's more, the advantage of the RBSA to BSA increases with the system load increasing.

\subsection{Effect of System Scale}

Changes in the system scale refer to the number of the schedulers and the number of the computing nodes, so here we also do two sets of experiments.

The first set of experiment is the effect of the number of the schedulers: we vary the number of the schedulers from $5$ to $20$, set the system load to $0.6$ and the number of computing nodes to $15$. The average task processing rate of each computing node is listed at Table \ref{table.4}.

\begin{center}
  \begin{table}
  \caption{Average Task Processing Rate of the Computing Nodes}
  \label{table.4}
  \begin{tabular}{|c|c|c|c|c|}
    \hline
    Computing node & 1-3 & 4-7 & 8-10 & 11 \\
    \hline
    Average Task Processing Rate & 0.01 & 0.02 & 0.033 & 0.06 \\
    \hline
    Computing node & 12 & 13 & 14 & 15 \\
    \hline
    Average Task Processing Rate & 0.05 & 0.03 & 0.025 & 0.03 \\
    \hline
  \end{tabular}
  \end{table}
\end{center}

The relative job arrival rates of $20$ schedulers are shown at Table \ref{table.5}.

  \begin{center}
  \begin{table}
  \caption{Relative Job Arrival Rate of Each Scheduler}
  \label{table.5}
  \begin{tabular}{|c|c|c|c|c|}
    \hline
    Scheduler & 1 & 2-5 & 6 & 7 \\
    \hline
    Relative Job Arrival Rate & 0.0035 & 0.01 & 0.006 & 0.005  \\
    \hline
    Scheduler  & 8 & 9-10 & 11 & 12 \\
    \hline
    Relative Job Arrival Rate &0.002 & 0.001 & 0.002 & 0.005 \\
    \hline
    Scheduler & 13 & 14 & 15 & 16 \\
    \hline
    Relative Job Arrival Rate & 0.003 & 0.0045 & 0.0037 & 0.0046 \\
    \hline
    Scheduler & 17 & 18 & 19 & 20 \\
    \hline
    Relative Job Arrival Rate & 0.0038 & 0.0063 & 0.0029 & 0.0048 \\
    \hline
  \end{tabular}
  \end{table}
  \end{center}

\begin{figure}
 \begin{center}
  \includegraphics[width=8cm,height=8cm]{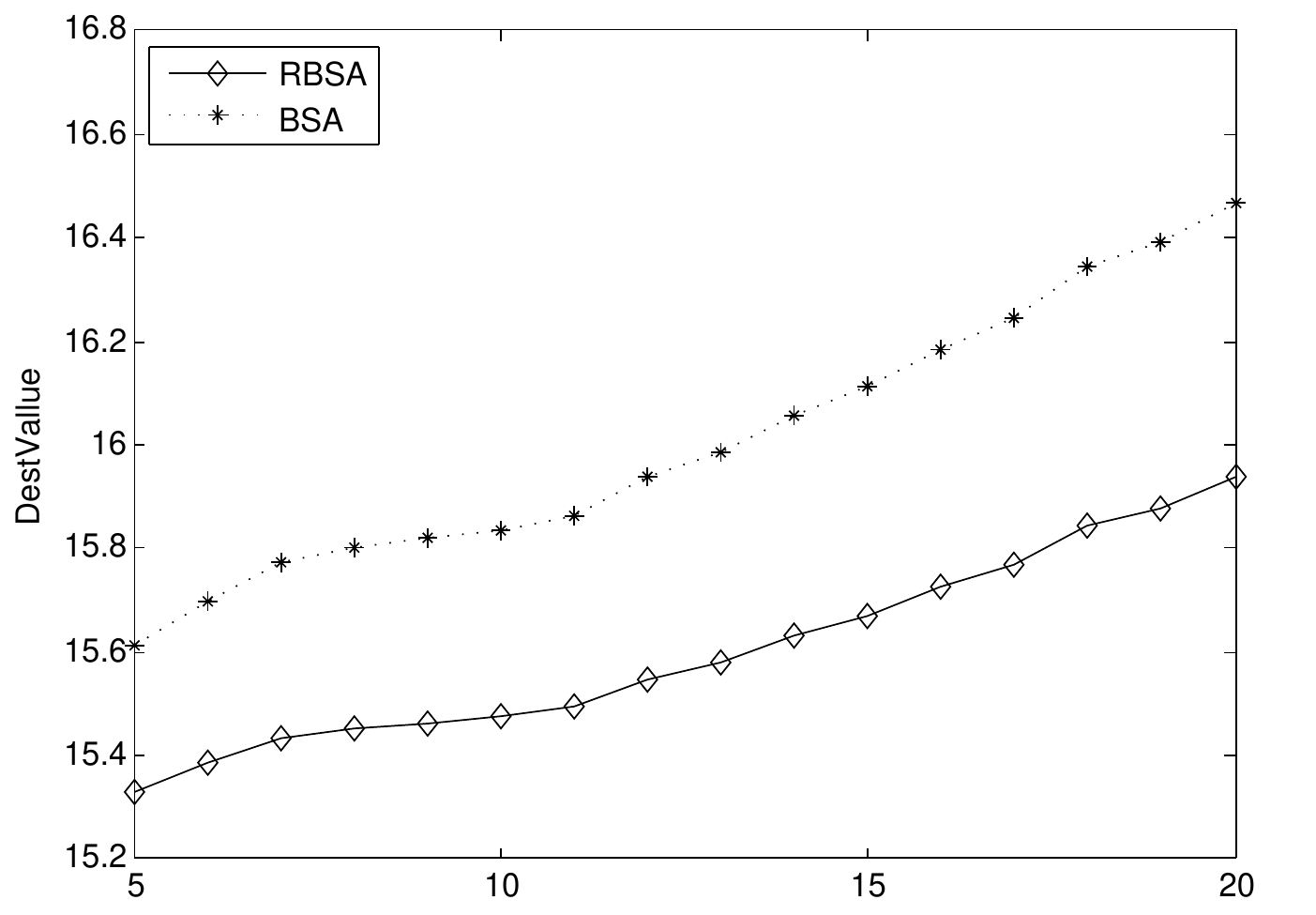}
 \caption{Objective Function Value Versus number of schedulers}
 \label{Fig.7}
 \end{center}
\end{figure}

Under the above initial condition, we can get the objective function values under different number of schedulers separately by RBSA and BSA as Fig.\ref{Fig.7} shows.

From Fig.\ref{Fig.7} we can see that: while the scheduler's number increases, the task scale increase and the objective function value increases, however ,the number of computing nodes and their average processing rates are not changed so that steady-state availability that the system provides decreases. What's more, the objective function value of the RBSA is always lower than that of the BSA and the advantage increases with the scheduler's number increasing.

The second set of experiment is the effect of the number of computing nodes. We vary the number of the computing nodes from $10$ to $20$, set the system load to $0.6$ and the number of schedulers to $15$. The average task processing rate of each computing node is listed at Table \ref{table.6}.

\begin{center}
  \begin{table}
  \caption{Average Task Processing Rate of the Computing Nodes}
  \label{table.6}
  \begin{tabular}{|c|c|c|c|c|}
    \hline
    Computing Node & 1-3 & 4-7 & 8-10 & 11\\
    \hline
    Average Task Processing Rate & 0.01 & 0.02 & 0.033 & 0.06 \\
    \hline
    Computing Node & 12 & 13 & 14 & 15 \\
    \hline
    Average Task Processing Rate & 0.05 & 0.03 & 0.025 & 0.03 \\
    \hline
    Computing Node & 16 & 17 & 18 & 19\\
    \hline
    Average Task Processing Rate & 0.025 & 0.033 & 0.028 & 0.025 \\
    \hline
    Computing Node & 20 & & & \\
    \hline
    Average Task Processing Rate &0.019 & & & \\
    \hline
  \end{tabular}
  \end{table}
\end{center}

The relative job arrival rates of $10$ schedulers are shown at Table \ref{table.7}.

\begin{center}
  \begin{table}
  \caption{Relative Job Arrival Rate of Each Scheduler}
  \label{table.7}
  \begin{tabular}{|c|c|c|c|}
    \hline
    Scheduler & 1 & 2-5 & 6 \\
    \hline
    Relative Job Arrival Rate & 0.0035 & 0.01 & 0.006 \\
    \hline
    Scheduler & 7 & 8 & 9-10 \\
    \hline
    Relative Job Arrival Rate & 0.005 & 0.002 & 0.001 \\
    \hline
  \end{tabular}
  \end{table}
\end{center}

\begin{figure}
 \begin{center}
  \includegraphics[width=8cm,height=8cm]{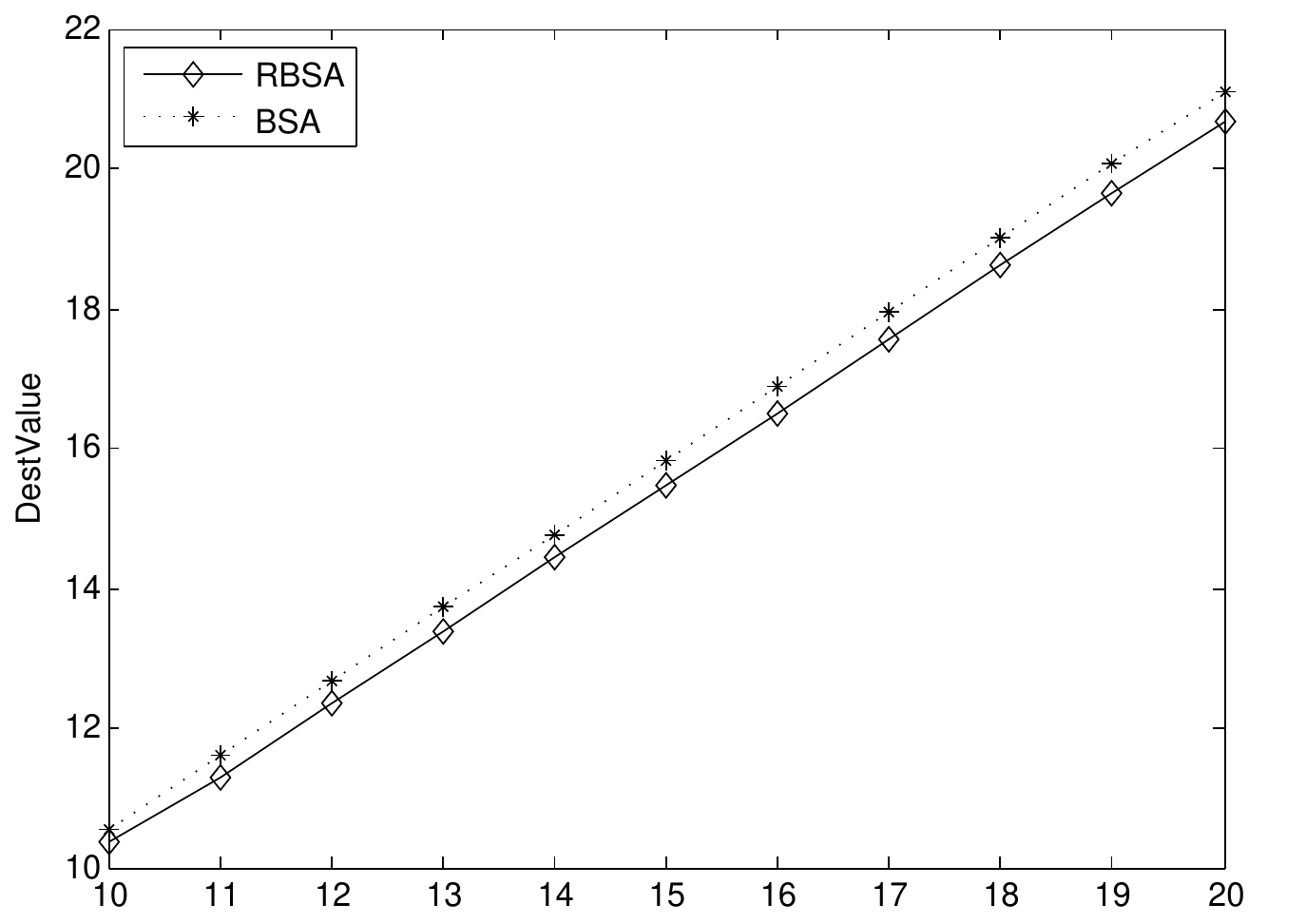}
 \caption{Objective Function Value Versus Number of Computing Nodes}
 \label{Fig.8}
 \end{center}
\end{figure}

Under the above initial condition, we can get the objective function values under different numbers of computing node separately by RBSA and BSA as Fig.\ref{Fig.8} shows.

From Fig.\ref{Fig.8}, we also can see that: the objective function of RBSA is always lower than that of BSA, namely, the RBSA is better than the BSA.

According to the above two sets of experiments, we can draw this conclusion : when the system scale increases, the advantage of the RBSA increases .It further illustrates the task scheduling algorithm this paper proposes can make the system provide greater steady-state availability.

\subsection{Fairness}

We measure fairness\cite{2} to users in the sense that users should receive the same level of utility. In this sense, fairness is achieved when the average steady-state availability for each user is the same. If one user has a higher average steady-state availability and another has a lower average steady-state availability, the task scheduling algorithm can be considered as unfair as it gives some users an advantage while it gives some other users a disadvantage.

A fairness index is given by

\begin{equation}
     FI=\frac{(\sum_{i=1}^nD_i)^2}{n\sum_{i=1}^nD_i^2}
\end{equation}
Where $D_{i}$ is the objective function value of scheduler $i$. When a scheduling algorithm's fairness index is closer to $1.0$, it is more fair. Here, we study the above three scheduling algorithm's fairness under different conditions such as system load, system size. Their initial conditions are the same as the above.

\begin{figure}
 \begin{center}
  \includegraphics[width=8cm,height=8cm]{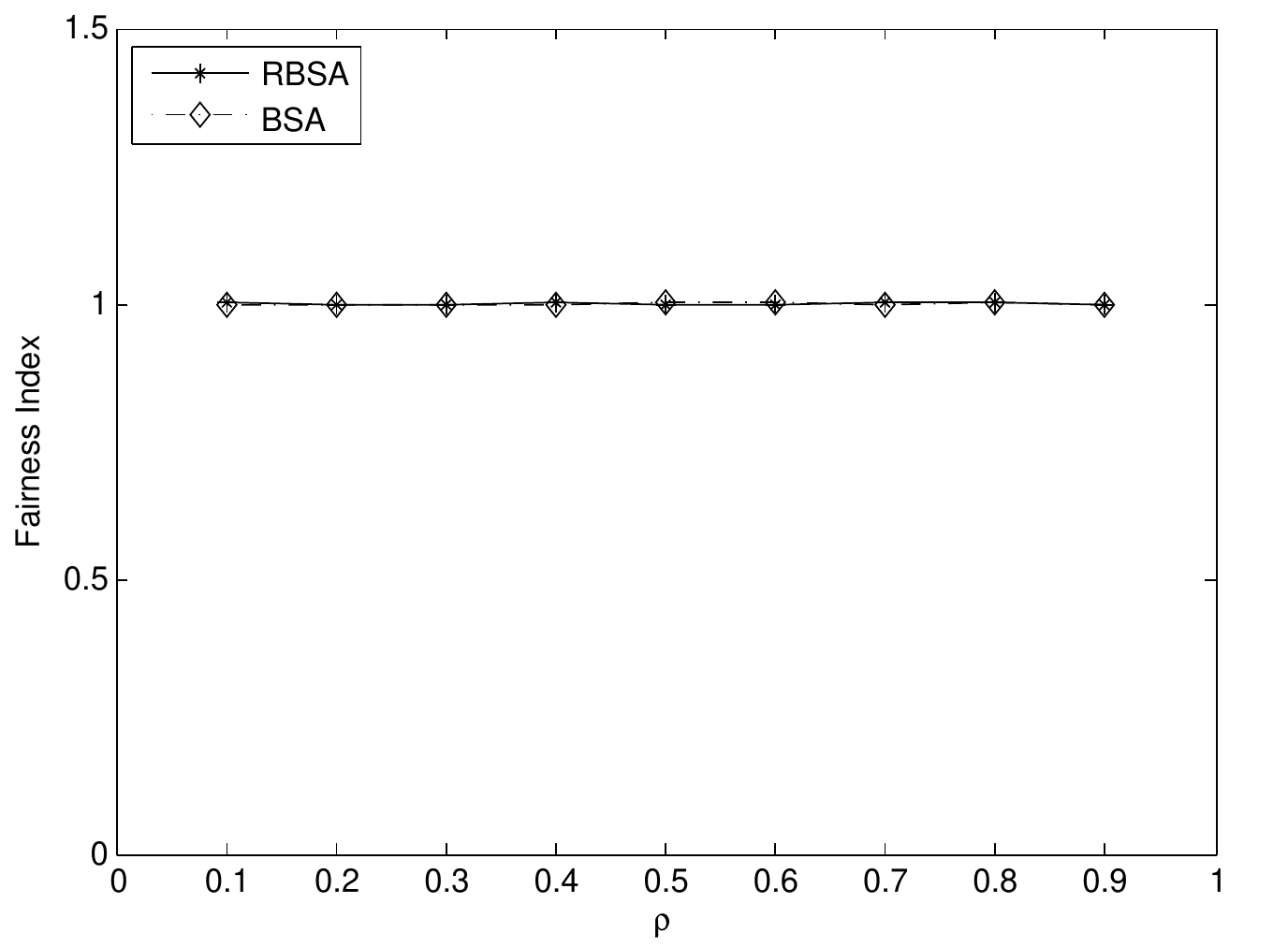}
 \caption{Fairness versus System Load}
 \label{Fig.9}
 \end{center}
\end{figure}

Fig.\ref{Fig.9} shows the fairness index when the system load varies from $0.1$ to $0.9$ under different algorithms. From this figure, we can find that the fairness indexes of both RBSA and BSA are equal to $1$, which means that both all algorithm are fair.

\begin{figure}
 \begin{center}
  \includegraphics[width=8cm,height=8cm]{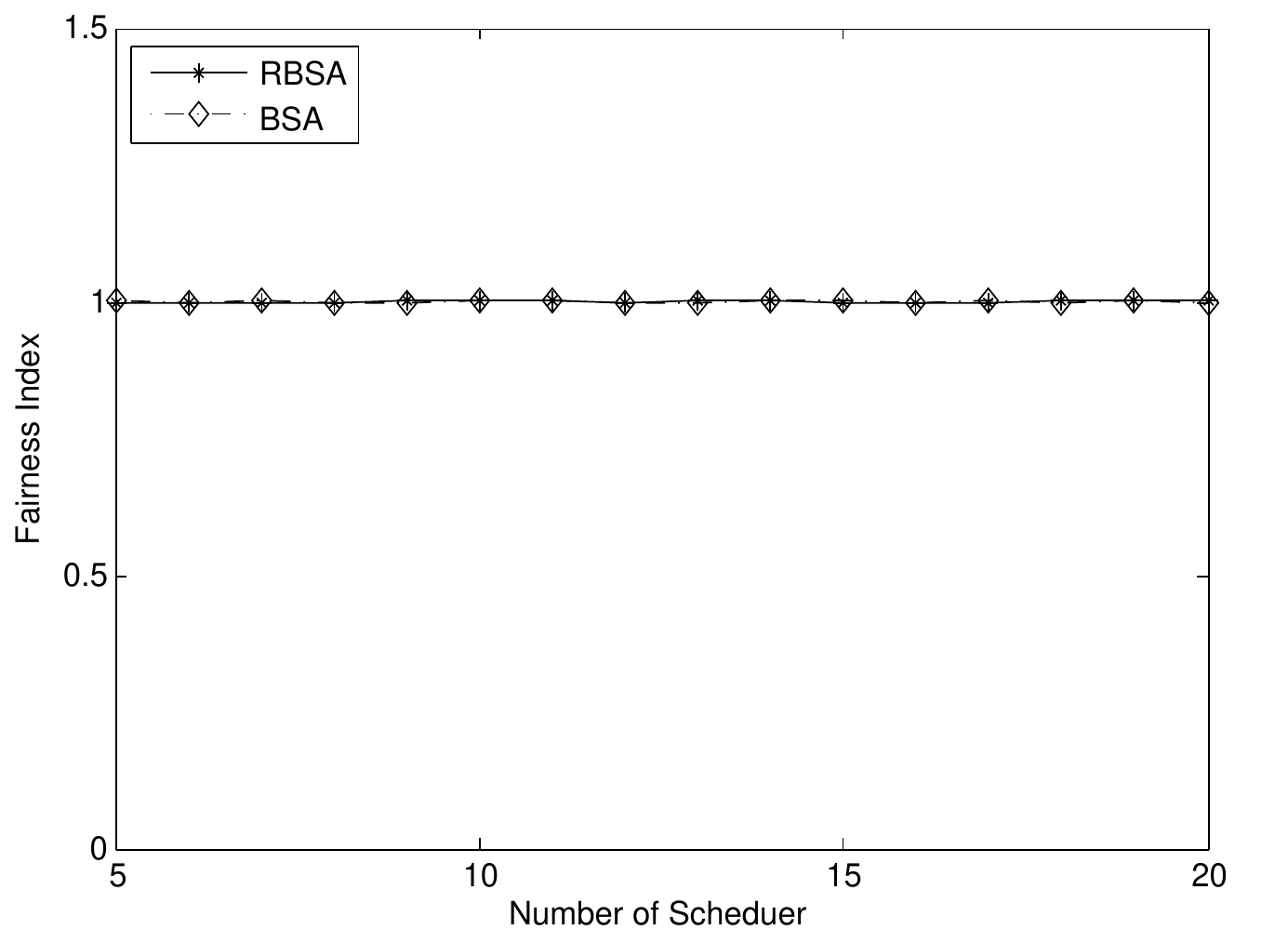}
 \caption{Fairness versus Number of Scheduler}
 \label{Fig.10}
 \end{center}
\end{figure}

\begin{figure}
 \begin{center}
  \includegraphics[width=8cm,height=8cm]{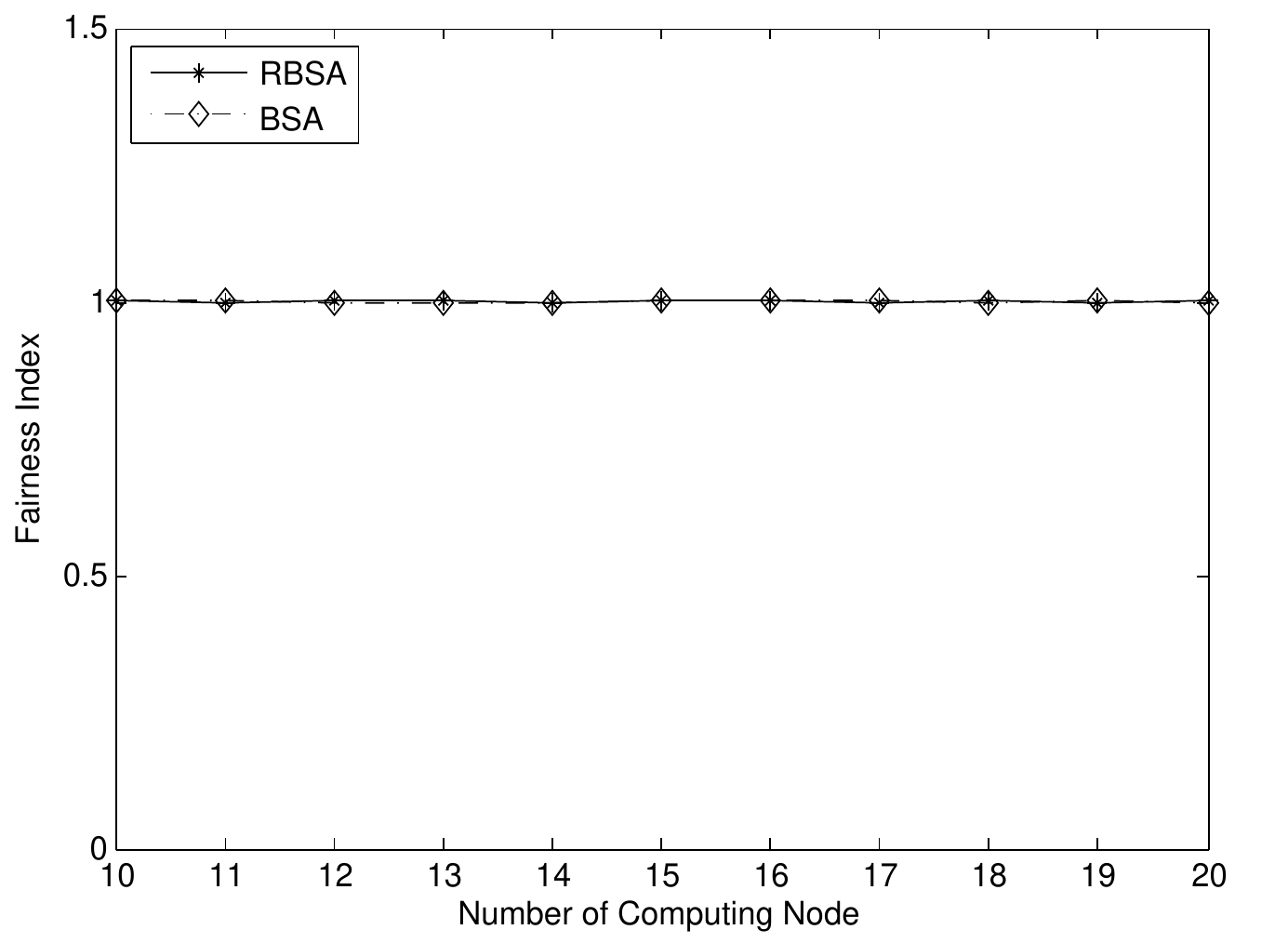}
 \caption{Fairness versus Number of Computing Node}
 \label{Fig.11}
 \end{center}
\end{figure}

When the number of scheduler changes, the fairness indexes of different algorithms are shown in Fig.\ref{Fig.10}. Fig.\ref{Fig.11} shows the fairness indexes when the number of computing node changes. From Fig.\ref{Fig.10} and Fig.\ref{Fig.11}, we can find that the fairness indexes are equal to 1.0 all the time for RBSA and BSA.

From the above three experiments , we can draw this conclusion: our proposed algorithm is fair.

\subsection{Convergence}

\begin{figure}
 \begin{center}
  \includegraphics[width=8cm,height=8cm]{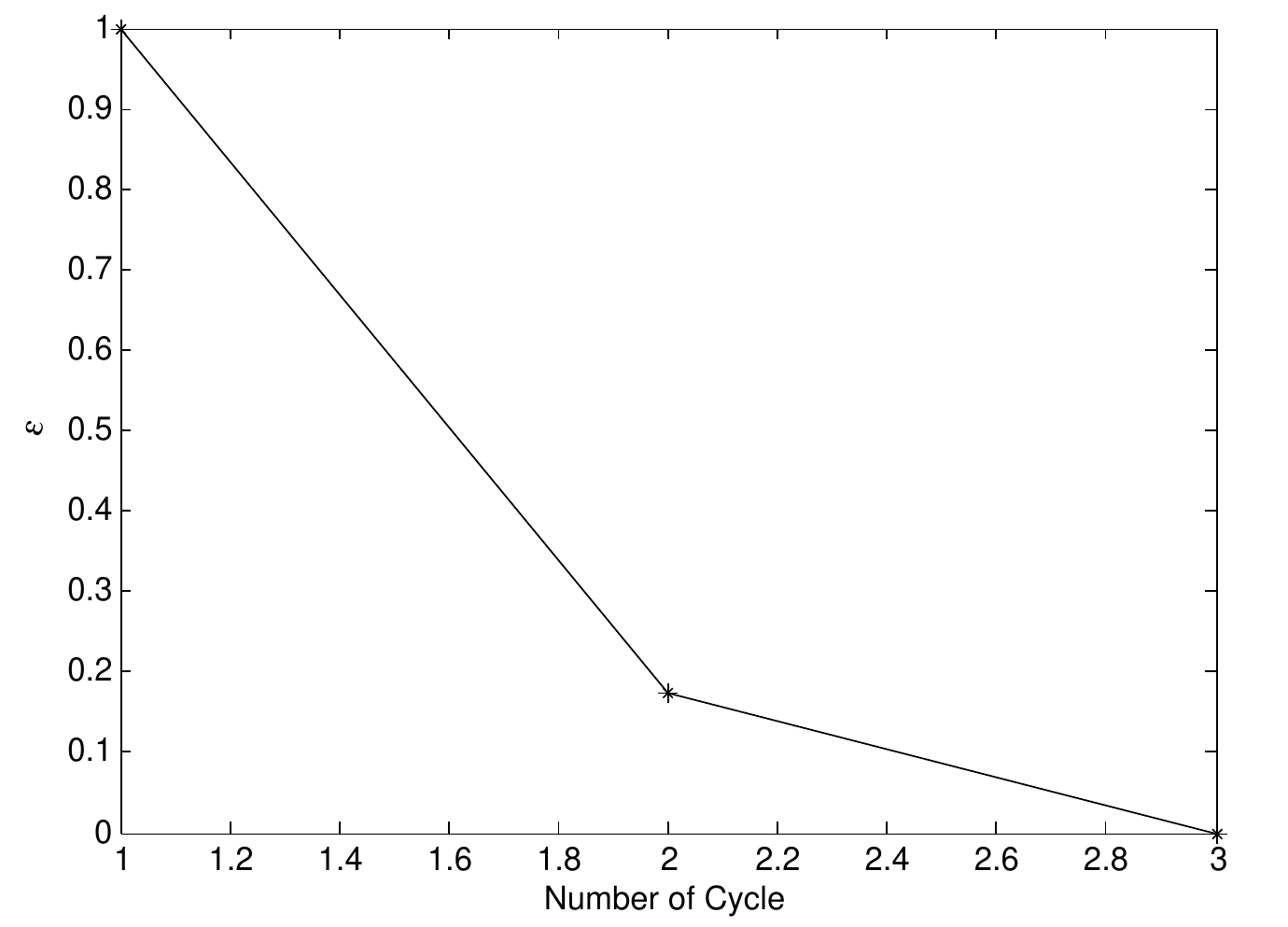}
 \caption{$\epsilon$'s Value while Cycle Number Increasing }
 \label{Fig.12}
 \end{center}
\end{figure}

When judging a task scheduling algorithm, we should also consider the merits of convergence of the algorithm. The speed of convergence directly affects the speed of implementation of the algorithm, thereby affecting the speed of the user to obtain results, which is essential to improve the user's experience. In this algorithm, the main factor affecting the speed of convergence is the speed of calculating allocation strategy $a$, which can be judged from $\varepsilon$. The value of $a$ will be the final value when $\varepsilon\leq 10^{-6}$. In order to clearly indicate $\epsilon$'s convergence speed, under the first experiment's initial condition in Section \ref{OFV}, we do a set of experiments to illustrate $\epsilon$'s values while calculating $a$, the result is shown in Fig.\ref{Fig.12}.

\begin{figure}
 \begin{center}
  \includegraphics[width=8cm,height=8cm]{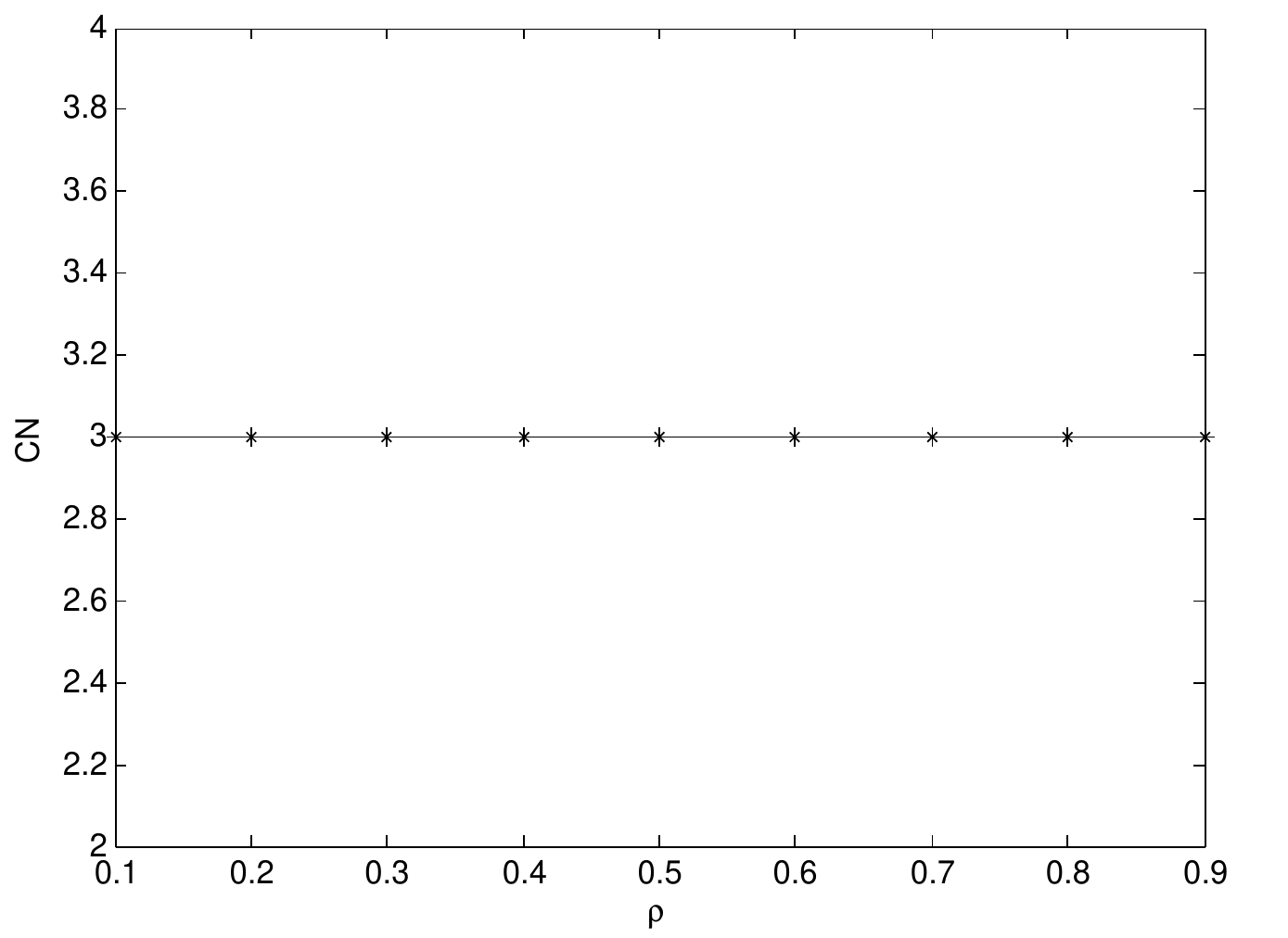}
 \caption{Cycle Number versus System Load}
 \label{Fig.13}
 \end{center}
\end{figure}

\begin{figure}
 \begin{center}
  \includegraphics[width=8cm,height=8cm]{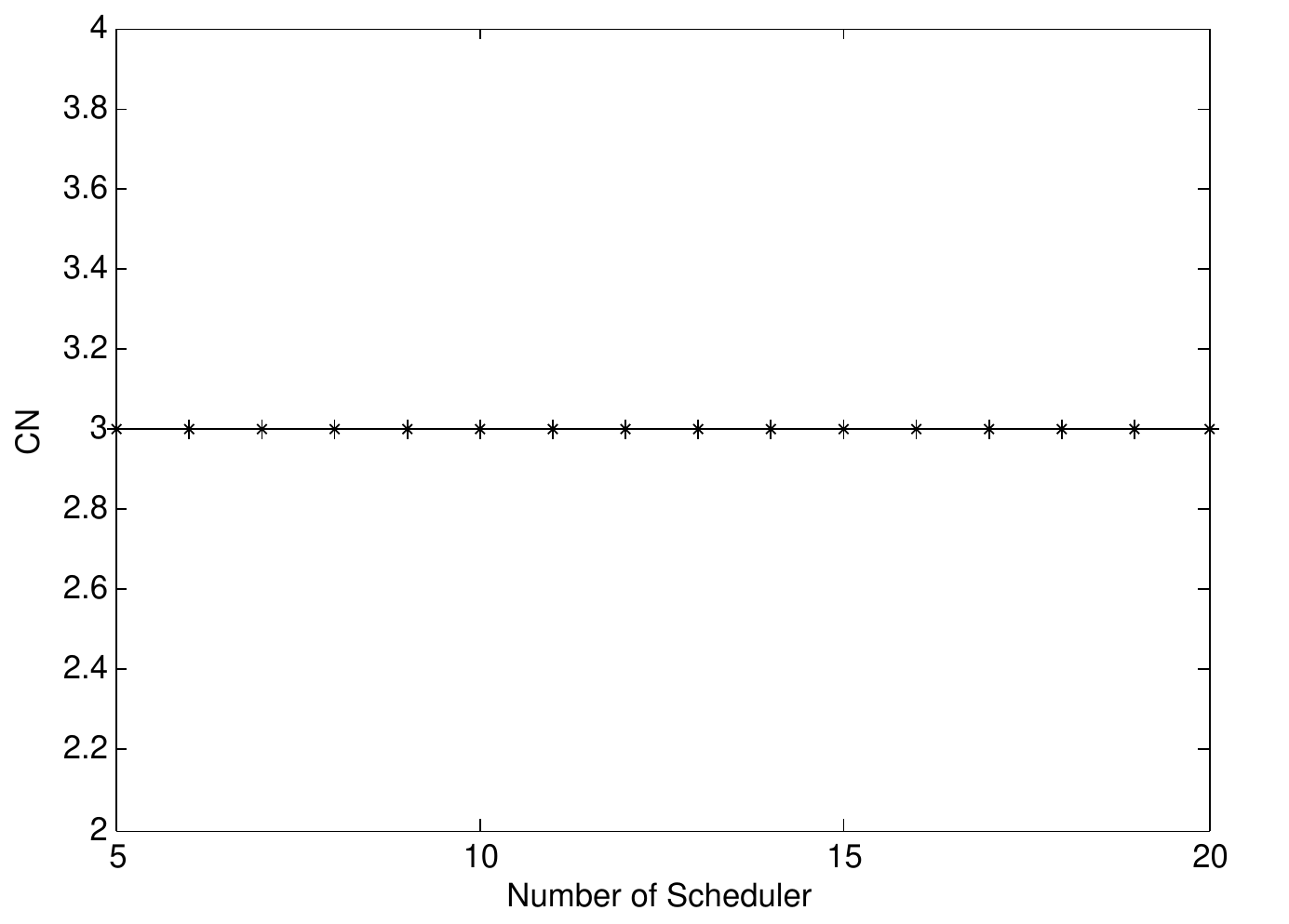}
 \caption{Cycle Number versus Number of Schedulers}
 \label{Fig.14}
 \end{center}
\end{figure}

\begin{figure}
 \begin{center}
  \includegraphics[width=8cm,height=8cm]{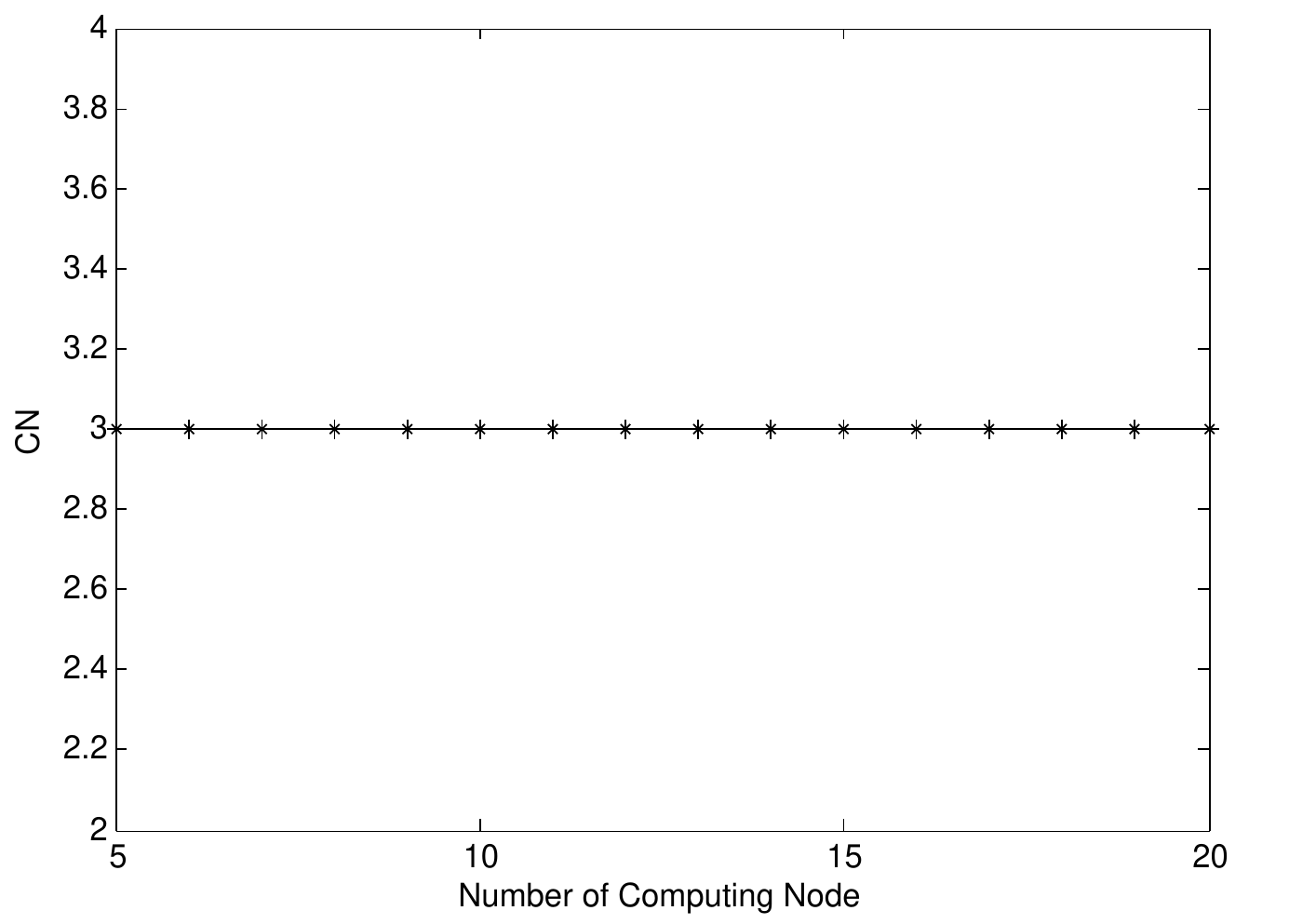}
 \caption{Cycle Number versus Number of Computing Nodes}
 \label{Fig.15}
 \end{center}
\end{figure}

From Fig.\ref{Fig.12}, we find that $\epsilon$'s value decreases very fast, it needs about $3$ cycle to get $a$'s final value. Then we study the speed of convergence under the condition of different system load and system scale. Corresponding to the initial condition of system load, schedulers' number, computing nodes' number as the above, we get three sets of results as Fig.\ref{Fig.13}, Fig.\ref{Fig.14}, Fig.\ref{Fig.15}, where X-coordinate respectively refers to system load $\rho$, schedulers' number , computing nodes' number, the y-coordinate represents Cycle Number which is labeled as CN.

As can be seen from Fig.\ref{Fig.13}, Fig.\ref{Fig.14}, Fig.\ref{Fig.15}, the speed of convergence is very fast under the above three conditions and this algorithm only needs about $3$ cycle to get $a$'s final value which clearly shows that the convergence speed of our algorithm is very well and very fast.

\section{Conclusions}

Aiming at the special requirements for reliability in cloud computing, this paper proposes a task scheduling algorithm based on steady-state availability and non-cooperative game. From the comparison with Balanced scheduling algorithm, we can find that this algorithm has a better steady-state availability under different system loads, and different numbers of computing nodes, and different numbers of schedulers. Moreover, it can ensure that every participant can have a fair opportunity to utilize the system resource.

However, this algorithm also has some deficiencies. It ignores the task's transmission reliability and the reliability that the scheduler processes tasks. Therefore, in the future, we should consider these conditions in order to make the algorithm more in line with the actual situation. In addition, instead of non-cooperative game theory, we are going to adopt cooperative game theory which may be better for the reason that the goal of this theory is to make the overall reliability best.

%
%

\label{lastpage}

\end{document}